\documentclass[%
twocolumn,
 amsmath,amssymb,
 aps,
 prl,
]{revtex4-1}

\usepackage{graphicx}
\usepackage{dcolumn}
\usepackage{bm}

\usepackage{color}
\definecolor{PMcolor}{rgb}{0,0,0}
\def\PM#1{\textcolor{PMcolor}{#1}} 
\definecolor{MHcolor}{rgb}{0,0,0}
 
\definecolor{EBcolor}{rgb}{0,0,0}

\begin{document}

\title{Acoustics of cubic bubbles: six coupled oscillators}

\author{Maxime Harazi}
\author{Matthieu Rupin}
\author{Olivier Stephan}
\author{Emmanuel Bossy}
\author{Philippe Marmottant}
\email{philippe.marmottant@univ-grenoble-alpes.fr}
\affiliation{University Grenoble Alpes, CNRS, LIPhy, F-38000 Grenoble, France}

\date{\today}

\begin{abstract}
In this manuscript we introduce cubic bubbles that are pinned to 3D printed millimetric frames immersed in water. Cubic bubbles are more stable over time and space than standard spherical bubbles, while still allowing large oscillations of their faces. We found that each face can be described as a harmonic oscillator coupled to the other ones.  These resonators are coupled by the gas inside the cube but also by acoustic interactions in the liquid.
We provide an analytical model and 3D numerical simulations predicting the resonance with a very good agreement. Acoustically, cubic bubbles prove to be good monopole sub-wavelength emitters, with non-emissive secondary surfaces modes.
\end{abstract}

\maketitle

\emph{Introduction.---}
Spherical air bubbles in water are known to be excellent acoustic resonators \cite{Minnaert1933,Leighton1994} because of the important compressibility of the gas compared to that of water, allowing large amplitude of vibration.
Furthermore, 
the wavelength at resonance is way bigger (500 times) than the bubble size. Bubbles are therefore good candidates for the design of new acoustic meta-materials including bubbles as sub-wavelength building blocks. Such meta-materials show an enhanced absorption at resonance \cite{Leroy2009,Bretagne2011} or feature remarkable properties such a negative index of refraction \cite{Brunet2014,Raffy2016,Lanoy2017} or the ability to focus acoustic energy with a sub-wavelength precision \cite{Lanoy2015}.

If large vibration amplitudes are possible in water, the experimentalist is quickly facing two important issues with spherical bubbles in water. First their stability in space is not insured because of buoyancy \cite{Heath1897}, meaning bubbles need to be kept under a surface or a net before acoustic excitation \cite{Leroy2005}. An other solution is to include bubbles in a gel \cite{Leroy2009}, but care as to be taken to choose a very soft gel and  not to add elastic effects. A second major challenge is the dissolution of gas in the liquid. Indeed the curvature of the surface is responsible for a  capillary overpressure  that triggers dissolution even into equilibrated or {oversaturated} water \cite{Epstein1950}. Stability analysis shows that a spherical bubble is thus always unstable in the long term.

Here we would like to introduce \emph{cubic bubbles}, designed in order to  overcome the issues of stability in position and in size, while still performing large amplitude of vibrations. Our approach is to pin the bubbles within 3D printed frames, resulting in bubbles with flat faces, and therefore no capillary overpressure.

\emph{Methods.---}
The 3D printed frames are crafted in photoresist polymer with a stereolithographic technology (Kudo3D,   $50\, \mathrm{\mu m}$ in horizontal resolution and $100\, \mathrm{\mu m}$ in vertical resolution). We chose the simplest shape for these frames: a cube (Fig. 1(a)).
The interior size of the cube is $2a \in [0.9-2.1 \, \mathrm{mm}] $ and the edge size $e=0.5 \, \mathrm{mm}$.
The frames are then silanized during 20 minutes using vapor-phase deposition of trichloro(perfluoro-octyle)silane. This renders the photoresist hydrophobic and  enhances the stability of the bubbles over time.

\begin{figure}[ht]
   \includegraphics[width=0.8\linewidth]{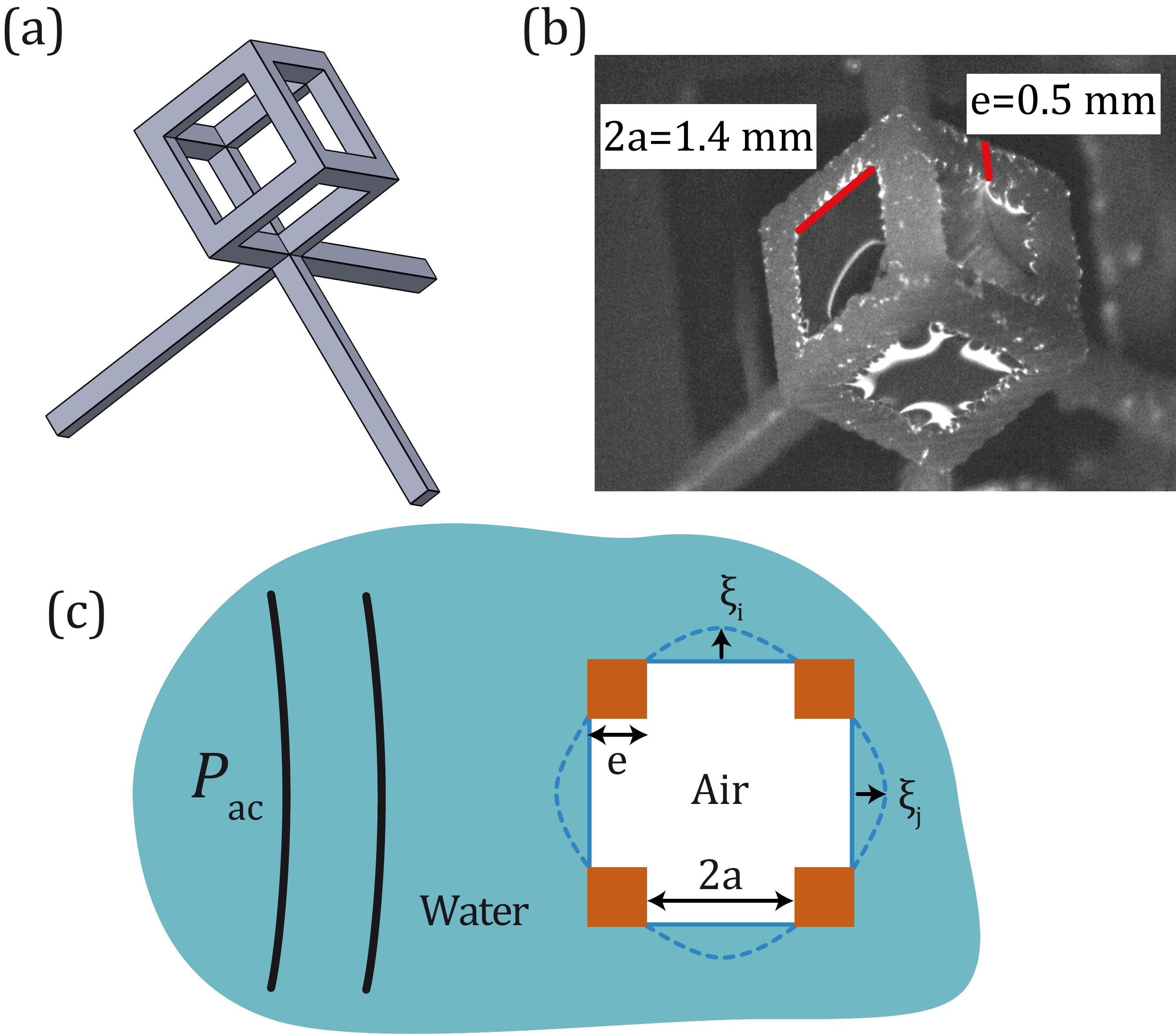}
\caption{(a) Design of a cubic frame (holding piedestal plate not represented)  (b) Top view of the immersion of the cubic frame in water: a gas bubble is trapped, and interfaces are visible on each faces  (c) Schematic drawing of a cross-section of the cube, showing the amplitude of vibration $\xi_i$ of an interface labeled $i$, while other faces $j$ have an amplitude $\xi_j$.}
    \label{fig1}
\end{figure}

When \PM{slowly} immersed in deionised water (equilibrated with the ambiant atmosphere), a volume of air is trapped in the frame, and six water-air interfaces are created (Fig. 1(b)).
\PM{The solid surface being hydrophobic, the interface usually attaches on the external corners, with a roughly   flat surface (blue continuous lines on figure 1(c)). Sometimes it happened that the interfaces attached on the the inner corners, these bubbles were discarded. Acoustic measurements are performed right after immersion.}
Bubbles are then excited in a tank ($30 \times 20 \times 20$ cm) using an underwater loudspeaker (Visaton, model FR13WP) driven by one period of sinusoid centered at $2.6\, \mathrm{kHz}$, {with a typical pressure of 100 Pa, low enough to drive only a linear response}. A hydrophone (Br\"uel \& Kj\ae r 8103) records the pressure signal $P(t)$ in presence of the bubble and $P^0(t)$ in the absence of the bubble, 
which allows by substraction to extract the signal emitted by the bubble only (Fig. 2(a)). The resonance frequencies of the bubbles were then extracted by computing the Fourier transform of the pressure signals and writing the relative bubble contribution 
\begin{equation}
   \hat{A}(f)=\frac{ \hat{P}(f) - \hat{P}^0(f)}{\hat{P}^0(f)}
\end{equation}
and looking for the frequency $f=f_\mathrm{res}$ such that $|\hat{A}(f)|$ is maximal 
(inset of Fig. 2(a)).

\emph{Stability.---} Having designed cubes with different lengths, we found that air bubbles are trapped if the opening length does not exceed  $2a_{max}= 2.1\,\mathrm{mm}$.
Above this length, water enters the frame. This is simply explained by the fact that at larger lengths the hydrostatic pressure induced by gravity overcomes surface tension $\sigma$ preventing water from entering.
The cross-over  occurs for a size of the order of the capillary length $l_c=(\sigma/\rho_l g)^{1/2}\simeq 2.7\, \mathrm{mm}$, which is indeed in the millimetric range of our observations. \PM{Because the} interfaces are pinned to the exterior of the edges after immersion (Fig. 1(b)), they are roughly flat \PM{cancelling any capillary over-pressure, contrary to spherical bubbles}. We observe that cubic bubbles can easily live more than few days in still water, without any temperature control or dissolved gas concentration control. \PM{After a few hours  days, we observe a slight tendancy for interfaces to enter in the frames, reducing the gas volume.}.  

\emph{Acoustic oscillations.---}
We  wanted to test if those bubbles were as good resonator as spherical bubbles, in spite of the rigid frame on which interfaces are pinned. Such a frame potentially restricts the oscillation amplitude.
Under pulse excitation, we observed that cubic bubbles behave as clear acoustic resonators, in the audible range (Fig. 2(a)), with a well defined resonance frequency and a quality factor $Q \simeq 20$, slightly lower than for spherical bubbles of similar size ($Q \simeq 35$) \cite{Leroy2004}. {The vibration amplitudes at the excitation pressure were minute and smaller than 10 micrometers, confirming a linear acoustic regime.}
 Experiments varying the aperture length $2a$ with a fixed pillar size $e$ show that this resonance frequency increases with smaller size {(Fig. \ref{fig2}(b))}.

A crude model can be made assuming the volume of gas {(inner space calculated using the cube dimensions $V_g=(2a)^3+6(2a)^2e$)} can be reshaped into an equivalent sphere of radius $R_{eq}$ given by {$V_g=4\pi R_{eq}^3/3$}. The Minnaert formula for resonance of spherical bubble \cite{Minnaert1933}, $f= \alpha /R_{eq}$ with $\alpha\simeq 3.24\,\mathrm{m/s}$, is then a first approximation, although it overestimates the experimental results. 

Here our goal is to present an analytical model that fully takes into account the geometry dictated by cube edges. Such a model should predict the effect of varying the number of open faces. We will present also numerical calculation of the propagation of sound through this complex geometry, to consolidate our analytical predictions.
We will see that the presence of the cubic frame modifies the frequency, and it is therefore not just a matter of volume of gas.

\begin{figure}[ht]
   \includegraphics[width=0.56\linewidth]{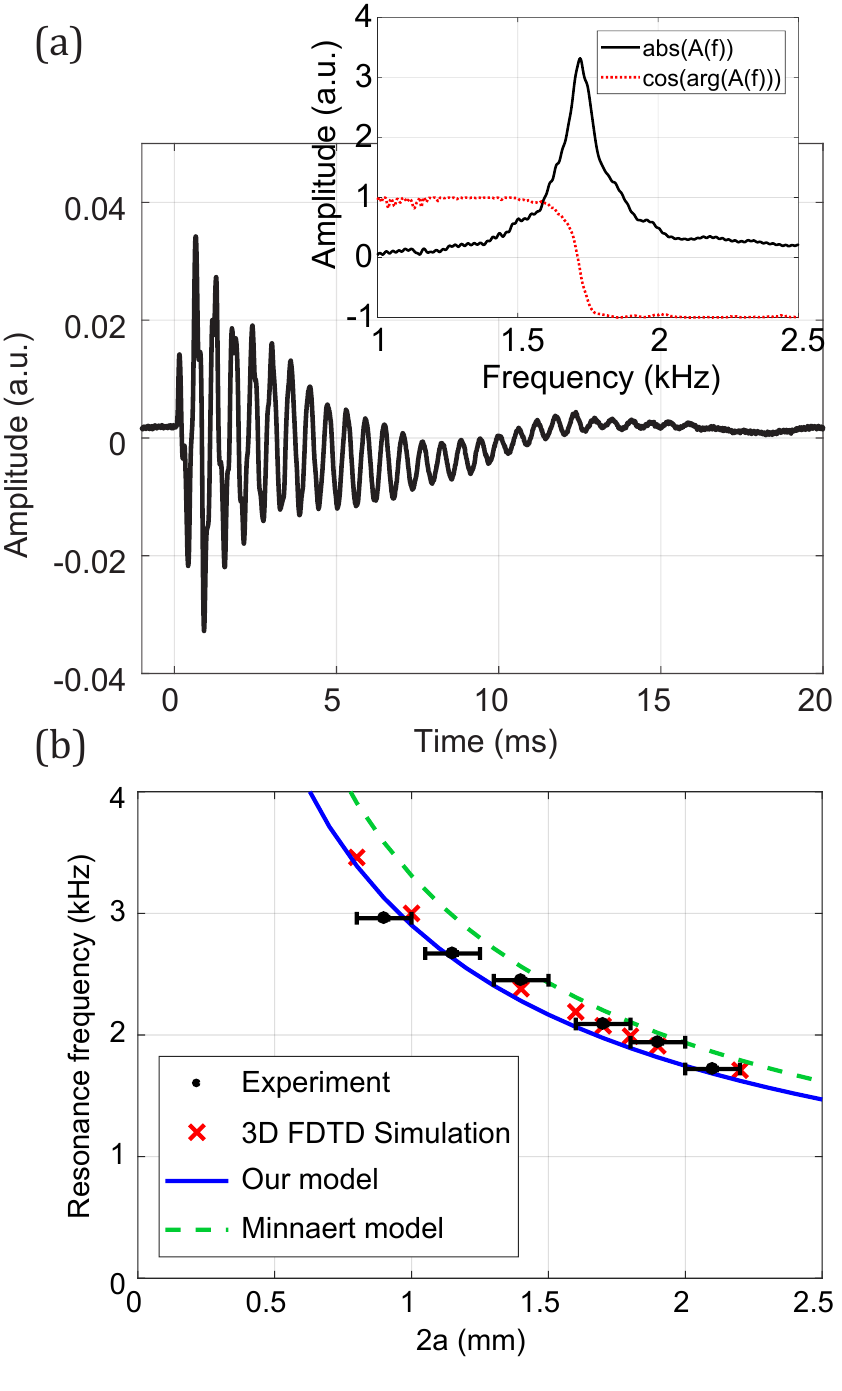}
   \caption{(a) Signal emitted by a bubble ($2a=2.1 \, \mathrm{mm}$)  after a pulse excitation. Inset: frequency spectrum of the signal emitted, normalized by the signal witout a bubble, showing the norm and the cosine of the phase.  
   (b) Measured frequency when varying the inner space between pillars $2a$, keeping $e=0.5\,\mathrm{mm}$ fixed. Crosses: 3D simulation. Line: theory for a cubic bubble with six coupled oscillators (Eq. \ref{eq:fresonance-coupled-interfaces}). Dashes: Minnaert frequency for a bubble of same gas volume. }
    \label{fig2}
\end{figure}

\emph{Model for a cubic bubble: 6 coupled oscillators.---}

Each of the 6 interfaces can be modeled as an individual oscillator, with a mass $m$ and stiffness $k$.
For this purpose we describe the displacement of interfaces {monitored by}
the displacement at the center of the interface  $\xi_i$ (see Fig. 1(c)), \PM{assumed to be small in front of the bubble size, $\xi \ll a$}. With an interface shape assumed to be parabolic {(and thus with an homogeneous curvature inducing an homogenous Laplace pressure jump)}, the average displacement over the surface is therefore $\overline\xi_i=\xi_i/2$.

According to \cite{Miller1983,Gelderblom2012} the reaction force necessary to set into motion the liquid just in front of the opening  is $m \ddot{\overline{\xi}_i}$ 
giving the added mass per interface  $m=\frac{32}{15}\rho_l a^3$.  Here we assume that the interface is circular of radius $a$, thus neglecting the corners, and that the aperture is embedded into an infinite plane.
In addition, the interface displacement gives rise to two elastic restoring forces. 
The first one is $-k_g{\overline{\xi}_i}$ a restoring stiffness due to the gas compression  $k_g=\kappa P_{0}(\pi a^{2})^{2}/V_g$, with $\kappa$ the polytropic exponent of the gas (close to adiabatic specific heat {ratio} for millimetric bubbles), $P_0$ the gas pressure, and $V_g$ the volume of gas trapped. Here we assumed the other interface are assumed to remain immobile.
The second force is $-k_\sigma {\overline{\xi}}_i$, with $k_{\sigma}=8\pi\sigma$  a capillary term due to the change in surface area. Tackling with millimetric dimensions, we find that $k_\sigma /k_g\sim 10^{-2}$ and capillary forces can be safely neglected, and the stiffness is simply $k=k_g+k_\sigma\simeq k_g$.

A single interface is therefore a mass-spring system with a resonance frequency $(k_g/m)^{1/2}/2\pi$. To verify this prediction, we designed a variation of our cubic bubble adding {solid walls instead of the openings, resulting in $N$ open faces with mobile interfaces and $6-N$ walls}. We find there is a very good agreement between this prediction and  the experiment with only one open face ({$N=1$,} first point of the curve in Fig. \ref{fig3}(a)).  

\begin{figure}[ht]
   \includegraphics[width=0.56\linewidth]{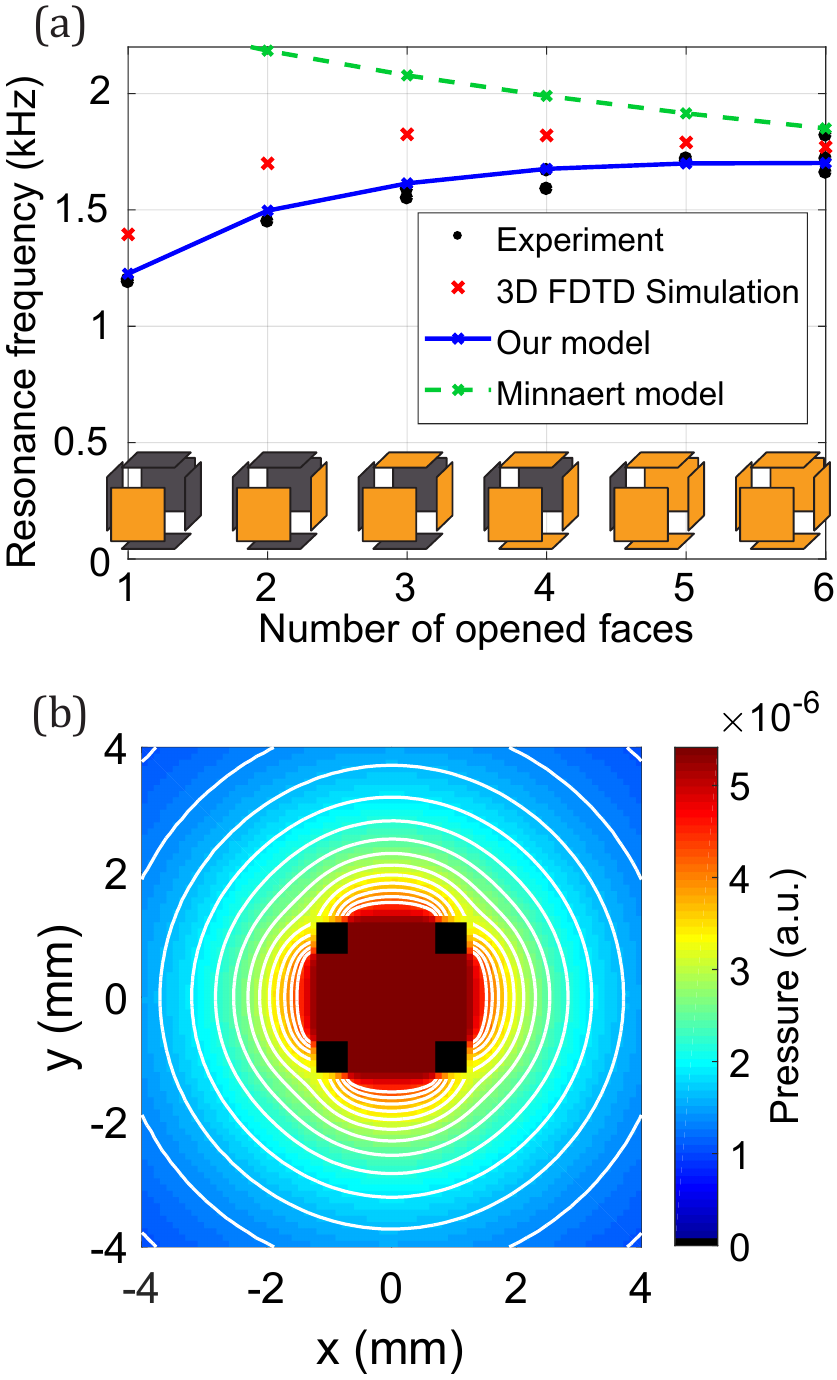}
   \caption{(a) Resonance frequency as a function of the number of opened faces (orange stands for open faces). Lines express the full theory (continuous line) and {Minnaert approximation}  (dashes), $2a=2.1$ mm. Crosses: 3D simulation. 
    (b) Pressure scattered by a cubic bubble \PM{($2a = 1.4$ mm)}: here an instantaneous cross-sectional snapshot   computed by 3D finite-difference time-domain  (FDTD) simulation, \PM{when the bubble oscillates at its natural resonance frequency after a wideband pulse excitation}.}
    \label{fig3}
\end{figure}

By opening an increasing number $N$ faces (chosen to be adjacent) we observe that the frequency depends strongly on $N$ (Fig. \ref{fig3}(a)). This suggests faces are not independent but coupled oscillators.

We found that these oscillators excited by the incoming acoustic pressure $P_\mathrm{ac}$ are actually coupled through the gas \emph{and} the liquid. The couplings are:
\begin{itemize}
\item a gas coupling, because when other faces $j$ also move they compress the common volume of gas, each face changing the gas pressure by $-k_g\overline\xi_j/\pi a^2$; 
\item an acoustic coupling through liquid, because other faces generate an oscillating flux $Q=\pi a^{2}\dot{\overline{\xi}_j}$ resulting in a  monopolar  acoustic pressure that decays with the distance to the center of the face $r$: $P{=\rho_l \dot{Q}/4\pi r}=\rho_l a^{2}\ddot{\overline{\xi}_j}/4r$ {(near field limit)}, modifying the pressure on other openings. The size being small compared to the wavelength we neglect retardation effects, and retain the far-field monopolar expression.
\end{itemize}

Overall, if we write the sum of forces acting on  {each interface labeled $i$ (displacing the effective mass $m$)}, we obtain the following set of coupled harmonic oscillators
\begin{equation}
    m\ddot{\overline{\xi}_i}
    +k_g\overline{\xi}_i=-\pi a^{2}P_\mathrm{ac}
    -\sum_{j\neq i}\left(
    k_g\overline{\xi}_j+
    \pi a^{2}\rho_l a^{2}\frac{1}{4r_{ij}}\ddot{\overline{\xi}_j}\right)
    \label{eq:model-coupled-oscillators}
\end{equation}
with 
$P_\mathrm{ac}$ applied acoustic pressure and $r_{ij}$ the distance between the centers of  faces $i$ and $j$.

Because of the  rotational symmetry of faces on a cube, all faces are equivalent, leading to a distribution of the distances $r_{ij}$ between one face $i$ and the five other ones independent of $i$.
The last term on the right hand side of equation Eq. \ref{eq:model-coupled-oscillators} is therefore independent of the face number $i$ for $N=6$ (with a restriction for $N<6$ that we will discuss just below).  Under the same applied pressure it is reasonable to consider that all faces oscillate with the same amplitude and phase: $\overline{\xi}_i=\overline{\xi}_j$. The previous equation becomes that of simple harmonic oscillator with an effective mass
\begin{equation}
    m_\mathrm{eff}=m+\pi\rho_l a^{3}\sum_{j\neq i}\frac{a}{4r_{ij}}
    \label{eq:mass-interactions}
\end{equation}
and an effective stiffness 
\begin{equation}
k_\mathrm{eff}=N k_g
\end{equation}
since all oscillators compress the same gas volume, demultiplying the resulting overpressure.

On a cube geometry the distances between the centers of adjacent faces is $r_{ij}=\sqrt{2}(a+e)$. \PM{Since wave travel in the liquid, we found it more appropriate to take the shortest path in liquid, that is $r_{ij}=2(a+e)$.}
If only $N$ faces are opened (chosen to be adjacent) it is straightforward to find by geometry that a good approximation of Eq. \ref{eq:mass-interactions} is $m_\mathrm{eff}=m+(N-1)\Delta m$ with $\Delta m=\pi\rho_l a^{3}\frac{a}{\PM{8}(a+e)}$.
This approximation is exact when $N=1,2,3$ adjacent faces  and a slight deviation occurs when $N=4,5,6$  because of the presence of an opposite face distant from $r_{ij}=\PM{4}(a+e)$ \PM{in terms of liquid path}. Note that strictly speaking this opposite face breaks the rotational symmetry for $N=4,5$\PM{, since not all faces have the same coupling terms (depending in $r_{ij}$) to the other ones, and the single oscillator model is not valid anymore}.

The resonance frequency of the coupled oscillators is $(k_\mathrm{eff}/m_\mathrm{eff})^{1/2}/2\pi$ and writes
\begin{equation}
f_\mathrm{res}=\frac{1}{2\pi}\sqrt{\frac{N\kappa P_{0}\pi^{2}a^{4}/V_g}{\frac{32}{15}\rho_l a^3+(N-1)\Delta m}}
\label{eq:fresonance-coupled-interfaces}
\end{equation}
Note the gas volume is given by the inner volume plus the space in between the pillars $V_g=(2a)^3+N(2a)^2 e$. 

Equation \ref{eq:fresonance-coupled-interfaces} is the main prediction of this letter: it {gives the resonance frequency} (Fig. \ref{fig2}b), with a better agreement for small $2a/e$ aperture ratios than the Minnaert prediction. For larger  $2a/e$ ratio the Minnaert prediction is a  better approximation, which is consistent with the fact that our model includes the mass $m$ for a vibration of an interface embedded in an infinite plane.

With this formula we can predict in addition the effect of a varying number of of open faces plotted on the experimental curve (Fig. \ref{fig3}(a), line), we find it works with no fitting parameter! 
{The agreement is excellent for a small number of openings, while the}
 Minnaert model is clearly out of scope for small $N$ numbers.
{Note the gas volume slightly increases with $N$ since solid walls are opened, explaining the decrease of the Minnaert prediction (Fig. \ref{fig3}(a), dashes).}

The present arrangement with $N=6$ provides a remarkable 
rotational symmetry that insures that all interfaces oscillators are equivalent \PM{(namely the system is symmetric when changing the face indices)}, contrary to a planar (or volumic) arrangement of monopole oscillators, since in the latter case oscillators on the edges are not equivalent to the ones in the center. Another difference with  separate wells in a planar arrangement \cite{Rathgen2007} is that the resonance frequency decreases with the number of wells with separate amount of gas, since only the effective mass is increased by acoustic coupling. Here the frequency initially increases with $N$ because the effective stiffness also increases with the gas coupling, the gas being shared by interfaces.

\emph{Monopole emission of a cubic bubble.---}
In order to specify the precise distribution of sound and verify the validity of our assumptions we conducted 3D numerical simulations of sound propagation through the cubic bubbles. The simulations were implemented with a finite-difference time-domain (FDTD) resolution of the elastodynamic equation for both fluid and solid structure, based on a freely available software developed in our group \cite{SimSonic}. We model the frame as a solid material (with properties close to that of plexiglas), and water and air were modeled as perfect fluids. A simulation volume of $20 \times 20 \times 20\ \mathrm{mm}^3$ was meshed with a grid step of $100\ \mathrm{\mu m}$, surrounded by perfectly-matched layers to mimick propagation in an unbounded medium. Wideband pressure pulses 
in the kilohertz range 
\PM{(2kHz center frequency, 200\% bandwidth) }
were propagated with and without the presence of the cubic bubble, analogous to the experimental situation, to derive the resonant frequency of the bubble. The validity of our approach was first confirmed by simulating the response of spherical air bubble of different sizes, which yielded values of the resonant frequency in excellent agreement with the value predicted by the Minnaert theory (accuracy better than $0.5\ \%$) (See further details in supplementary information).
\\ 
Our simulations confirmed the behaviour observed experimentally for the effect of size (Fig. \ref{fig2}(b)) and number of faces (Fig. \ref{fig3}(a)), while shedding light on the determinant parameters of the resonant frequency: it was observed by removing the solid frame (only possible with simulations), the resonant frequency of a cubic air bubble is very close (typically less than $1\ \%$ relative difference)  to that of a spherical bubble of equivalent volume. It is therefore the presence of the solid frame which turns the cubic bubble into six coupled oscillators each pinned to a rigid frame, as assumed by our analytical model. Another important information from the simulations is that cubic bubbles behave as monopole acoustic sources (Fig. 3(b)) as for spherical bubbles.

\emph{A side effect: capillary surface modes.---} 
At much larger amplitudes of acoustic pressure, we observed resonant surface modes with nodes and antinodes across the square interface (see Fig. \ref{fig4a}, red arrow).

In order to reach such higher acoustic pressures we designed a specific set-up: a steel tank with glass windows (see Supplementary Fig. 1), with
a shaker pushing/pulling directly the water though a tight orifice. 
A precise map of the vibration amplitude those surface modes was performed on a customized cube (see Supplementary Fig. 2).  Such an elevation map is morphed  on the top face of the image Fig. \ref{fig4a}. 

We interpret those modes as standing capillary waves. The modes are probably directly excited by the contact line on the edges, without any parametric instability. Note these modes are not emissive: there is no signature of them  in the frequency spectrum from hydrophone recordings.




\begin{figure}[ht]
   \includegraphics[width=0.56\linewidth]{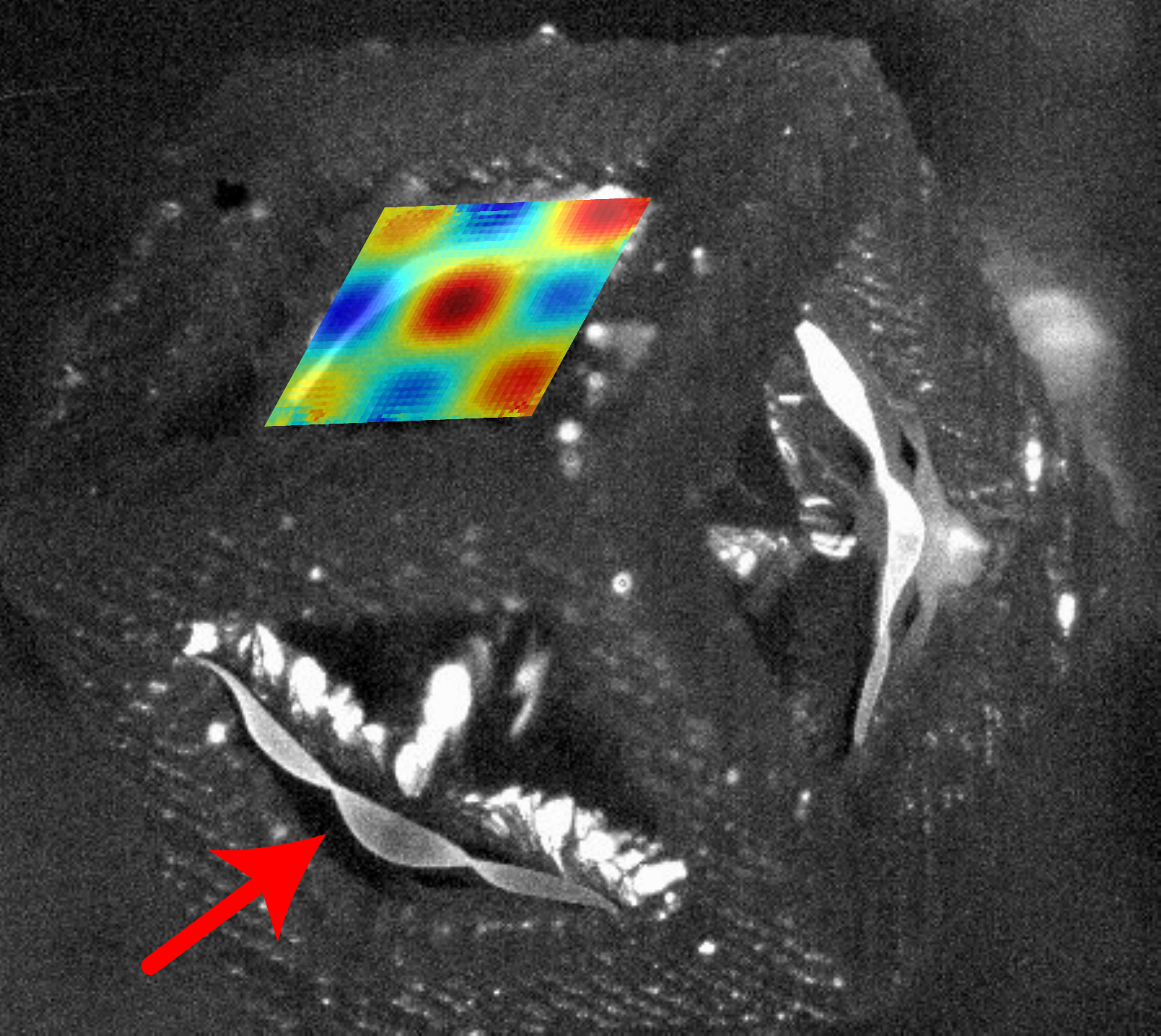}
   \caption{Surface modes at $f=760 \, \mathrm{Hz}$, for a cube of inner size $2a=1.7\,\mathrm{mm}$ The red arrow shows the vibration of the reflection of line of light. The image of a front measurement of the vibration amplitude performed on a single face is morphed and superimposed on the top face the cube. It shows a standing wave with $3 \times 3$ half-wavelengths. \PM{Here, the interfaces were slightly pushed inside when closing the water-tight tank.} See also \PM{the} supplementary movie showing the vibration mode.}
    \label{fig4a}
\end{figure}

\emph{Conclusion and perspectives.---}
We have shown that cubic bubbles are good candidates for the design of acoustic meta-materials: they are sub-wavelength resonators, with a good quality factor and an isotropic emission field close to the one of spherical bubbles. Furthermore they are easy to assemble in a large number using 3D stereolithography, allowing for stability in space and time. \PM{Because they have flat interfaces, there is no capillary over-pressure that would tend to speed up dissolution. Moreover there is no rectified diffusion under oscillations that would inflate the gas volume, as is the case with  spherical bubbles \cite{Fyrillas1994,Lauterborn2010}}. Future work will aim at understanding the interaction between a large number of cubic bubbles, paving the way to precisely designed acoustic meta-materials. 

\begin{acknowledgements} 
P.M. acknowledges financial support from the European Community's Seventh Framework Programme (FP7/2007-2013) ERC Grant Agreement Bubbleboost no. 614655 \PM{and an anonymous referee for the suggestion to consider the liquid path to model the interaction between faces}. M.H. acknowledges L\'eo Denat for support in the experiments, in addition to Benjamin Dollet, Thomas Combriat, Patrice Ballet and Thibaut Metivet for useful discussions.
\end{acknowledgements}


\clearpage
\begin{titlepage}
\maketitle
\thispagestyle{empty}
\end{titlepage}
\section*{Supplementary material for the mansucript 'Acoustics of cubic bubbles: six coupled oscillators'}
\subsection{Supplementary figures}
\begin{figure}[ht]
   \includegraphics[width=\linewidth]{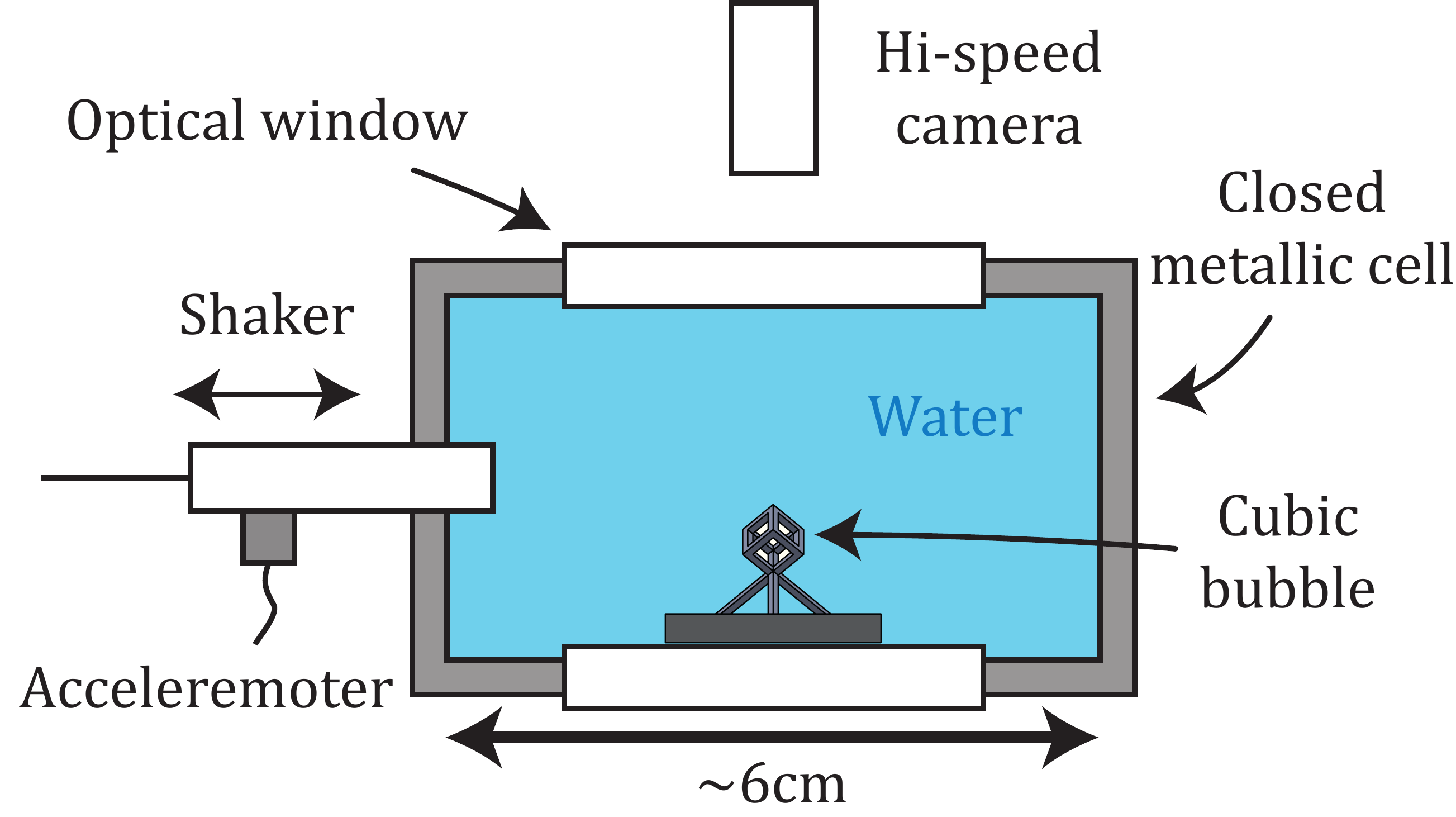}\\[3mm]
   \noindent{SUPPLEMENTARY FIGURE 1: Experimental set-up used for the generation of high pressure amplitudes. The bubble is confined in a closed metallic cell filled with water. A shaker pushes directly on the water, compressing directly the bubble. An hydrophone and an high-speed camera allow respectively for monitoring of the applied force and displacement of the bubble interface(s).}
    \label{figSup_cell}
\end{figure}

\begin{figure}[ht]
   \includegraphics[width=\linewidth]{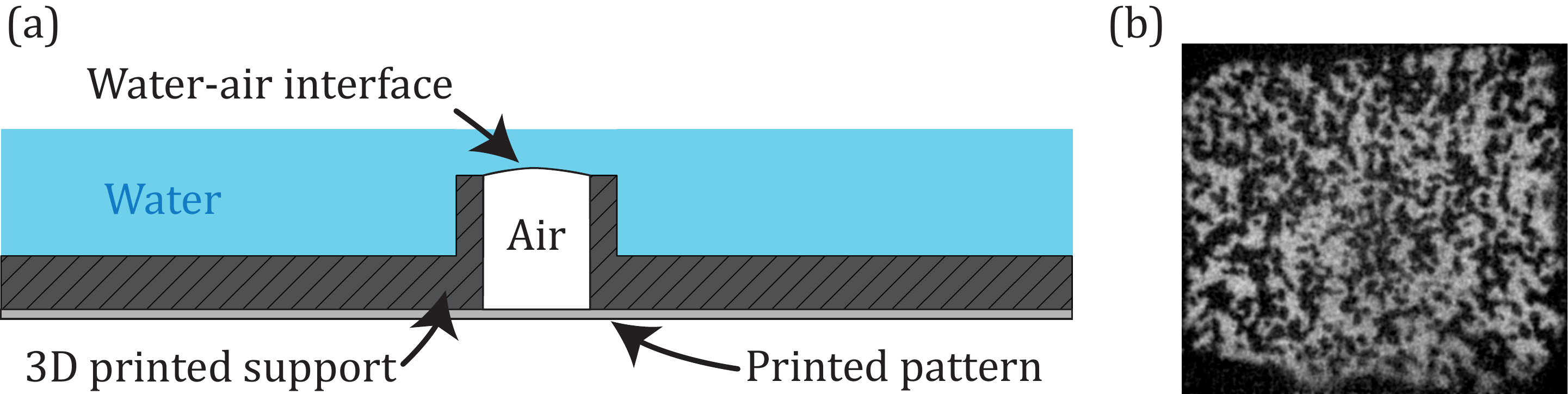}\\[3mm]
   \noindent{SUPPLEMENTARY FIGURE 2: (a) Configuration used for visualization of the capillary modes: a random pattern is printed on a transparent sheet and glued to a cube with two opposite openings (other faces are closed). (b) Visualization of the random pattern through the bubble. Its deformation due to the interface displacement allows for reconstruction of the height of the interface \textit{via} digital image correlation (DIC \cite{blaber_ncorr:_2015}) and double integration \cite{moisy_synthetic_2009}.}
    \label{figSup_capillary}
\end{figure}

\subsection{Supplementary movie}
Movie.avi : Oscillations of the bubble faces (top view, only three faces
are visible) under oscillation of the shaker (high amplitude) at 761 Hz,
corresponding to a $3 \lambda/2$ capillary mode. Note that the additional
displacement map visible on figure 4 of the main manuscript was acquired via a different setup (see supplementary figure 1), but at the same frequency. \\[3mm]

\subsection{Supplementary information regarding the 3D numerical simulations.---}
As introduced briefly in the main text, we perfomed numerical simulations of ultrasound propagation through cubic bubbles via a finite-difference time-domain (FDTD) approach. The numerical scheme implemented in this work is based on a leap-frog finite-difference resolution of the elastodynamic equations as first proposed by Virieux in a seminal work in the field of geophysics \cite{virieux1984sh}. We used a freely available FDTD software developped in our group \cite{SimSonic} , initially developed to model ultrasound propagation in bone and soft tissue \cite{bossy2004three}, as the Virieux scheme is known to accuraterly model fluid-solid interaction. In practice, the propagation medium is simply described by the list of the physical parameters (density, elastic constants) at each point of the simulation grid. The spatial grid step was set to $100\ \mu m$, fine enough to accurately model the geometry of the cubic bubble and the solid frame (0.5 mm in thickness) and much smaller than the smallest wavelength in the kHz range (typical at least several cm in air). The time step was imposed by the CFL (Courant-Friedrichs-Lewy) stability condition to about 20 ns, much smaller than the time scale of the ultrasound pulses. The thickness of the perfectly-matched layers surrounding the simulation domain was 10 mm (100 grid points), which was observed to be sufficient to avoid spurious reflections from the grid boundaries. Each simulation typically required 5GB of RAM, and computation times of the order of a couple of days for 10 ms of ultrasound propagation when ran in parallel on 32 CPUs. All the configuration files for the simulations of this work are available on demand and can be used to readily reproduce our results with our freely available software\cite{SimSonic}.\\
To simulate the cubic bubble, three different sets of physical parameters were associated to grid points, corresponding to water (surrounding medium), air (bubble) or solid (frame). As a key feature of numerical simulations, the properties of the cubic frame could be straightforwardly changed from one type of material to another, or made perfectly rigid. In particular, we observed no significant difference in the resonant frequency when the frame was modeled as plexiglas or as a perfectly rigid structure. Moreover, if the frame was modelled as water (equivalent to having no frame while keeping the cubic shape of the air bubble), we observed that the resonant frequency of the cubic bubble was very close to that of a spherical bubble of equivalent volume, only slightly larger by typically $1\ \%$. On the other hand, the effect of the solid frame is to reduce the resonant frequency by typically $5\ \%$. The larger relative reduction (about $10\ \%$ predicted the analytical model is likely to due to an overestimation of the effect of the frame rigidity which is modelled as infinite baffle for each oscillator whereas it is in practice of finite dimension.

\end{document}